\documentclass{aa}
\usepackage{graphicx}
\usepackage{natbib}
\usepackage{helvet}
\usepackage[utopia]{mathdesign}
\usepackage{microtype}
\usepackage{hyperref}
\hypersetup{
    colorlinks,%
    citecolor=blue,%
    linkcolor=blue,%
    urlcolor=blue
}

\usepackage{balance}

\usepackage{tikz}
\usetikzlibrary{calc}
\usetikzlibrary{through}
\usetikzlibrary{intersections}
\usetikzlibrary{decorations.pathmorphing}
\usetikzlibrary{decorations.text}

\def\msol{\ensuremath{\mathrm{M}_\odot}}
\def\msolyr{\ensuremath{\mathrm{M}_\odot\,\mathrm{yr}^{-1}}}
\def\lsd{\ensuremath{L_\mathrm{sd}}}
\def\kms{\ensuremath{\mathrm{km\,s^{-1}}}}
\def\ergs{\ensuremath{\mathrm{erg\,s^{-1}}}}
\def\ergcms{\ensuremath{\mathrm{erg\,cm^{-2}\,s^{-1}}}}

\newcommand{\refsec}[1]{Sect.~\ref{sec:#1}}
\newcommand{\refeq}[1]{Eq.~(\ref{eq:#1})}
\newcommand{\reffig}[1]{Fig.~\ref{fig:#1}}

\begin{document}
   \title{Unraveling the high-energy emission components\\ 
       of gamma-ray binaries}
   \titlerunning{GeV-TeV emission model for LS\,5039}

   \author{V.~Zabalza \inst{1} \and V.~Bosch-Ramon \inst{2} \and F.~Aharonian
        \inst{3,1} \and D.~Khangulyan \inst{4} }

   \institute{
        Max-Planck-Institut f\"ur Kernphysik, Saupfercheckweg 1, Heidelberg
        69117, Germany, \\ \email{Victor.Zabalza@mpi-hd.mpg.de} 
        \and 
        Departament d'Astronomia i Meteorologia, Institut de Ci\`encies
        del Cosmos (ICC), Universitat de Barcelona (IEEC-UB),
        Mart\'i i Franqu\`es, 1, E08028, Barcelona, Spain
        \and
        Dublin Institute for Advanced Studies, 31 Fitzwilliam Place, Dublin 2,
        Ireland \and 
        Institute of Space and Astronautical Science/JAXA, 3-1-1 Yoshinodai,
        Chuo-ku, Sagamihara, Kanagawa 252-5210, Japan
             }

   \date{Received 18 October 2012 / Accepted 24 December 2012}
 
  \abstract
   {
       The high and very high energy spectrum of gamma-ray binaries has become a
       challenge for all theoretical explanations since the detection of
       powerful, persistent GeV emission from LS\,5039 and LS\,I\,+61\,303 by
       \emph{Fermi}/LAT. The spectral cutoff at a few GeV
       indicates that the GeV component and the fainter, hard TeV emission above
       100\,GeV are not directly related.
   }
   {
       We explore the possible origins of these two emission components in the
       framework of a young, non-accreting pulsar orbiting the massive star,
       and initiating the non-thermal emission through the interaction of the stellar
       and pulsar winds.
   }
   {
       The pulsar/stellar wind interaction in a compact-orbit binary gives rise
       to two potential locations for particle acceleration: the shocks at the
       head-on collision of the winds and the termination shock caused by
       Coriolis forces on scales larger than the binary separation. We explore
       the suitability of these two locations to host the GeV and TeV emitters,
       respectively, through the study of their non-thermal emission along the
       orbit. We focus on the application of this model to LS\,5039 given its
       well-determined stellar wind with respect to other gamma-ray binaries.
   }
   {
       The application of the proposed model to LS\,5039 indicates that these two
       potential emitter locations provide the necessary conditions for
       reproduction of the two-component high-energy gamma-ray spectrum of LS\,5039.
       In addition, the ambient postshock conditions required at each of the
       locations are consistent with recent hydrodynamical simulations.
   }
   {
       The scenario based on the interaction of the stellar and pulsar
       winds is compatible with the GeV and TeV emission observed from
       gamma-ray binaries with unknown compact objects, such as LS\,5039 and LS\,I\,+61\,303.
   }

   \keywords{radiation mechanisms: non-thermal --- stars: individual (LS 5039) --- gamma rays: theory}

   \maketitle
%

\section{Introduction}\label{sec:intro}

Gamma-ray binaries are compact binary systems composed of a massive star and a
compact object, either a black hole or a neutron star. They exhibit non-thermal
emission from radio to high-energy (HE, $>$100\,MeV) and very high energy (VHE,
$>$100\,GeV) gamma rays, which is typically modulated with the orbital period,
and most of their energy output is emitted in the MeV to TeV range.  Four
gamma-ray binaries have been confirmed to emit at VHE: \object{LS\,5039}
\citep{2005Sci...309..746A}, \object{PSR\,B1259$-$63}
\citep{2005A&A...442....1A}, \object{LS\,I\,+61\,303}
\citep{2006Sci...312.1771A}, and \object{HESS\,J0632$+$057}
\citep{2009MNRAS.399..317S}.  The \emph{Fermi}/LAT source
\object{1FGL\,J1018.6$-$5856} \citep{2012Sci...335..189F} appears to have a
binary origin as well, but its candidate VHE counterpart is still not clearly
associated \citep{2012A&A...541A...5H}.  PSR\,B1259$-$63 is known to host a
young, non-accreting pulsar \citep{1992ApJ...387L..37J}, but the nature of the
compact objects in all other gamma-ray binaries remains unknown. Flaring HE or
VHE gamma-ray emission has also been detected from the X-ray binaries Cyg~X-1
\citep{2007ApJ...665L..51A} and Cyg~X-3 \citep{2009Sci...326.1512F,
2009Natur.462..620T}, but this emission is neither dominant nor persistent as is
the case for the gamma-ray binaries mentioned above.

Two different scenarios have been extensively proposed in the literature to
account for the non-thermal emission from LS\,I\,+61\,303 and LS\,5039, probably
the most-studied gamma-ray binaries along with PSR\,B1259$-$63: a microquasar
scenario, with relativistic jet formation through matter accretion onto the
compact object \citep[see, e.g.,][for a review]{2009IJMPD..18..347B}, and the
pulsar wind shock (PWS) scenario, with particle acceleration being the result of
the interaction between the pulsar and stellar winds, first proposed by
\cite{1981MNRAS.194P...1M}. LS\,5039 is the best gamma-ray binary to test wind
interaction models because of its radial, well-determined stellar wind, as
compared to LS\,I\,+61\,303, PSR\,B1259$-$63 and HESS\,J0632+057, which all
contain Be stars. We will therefore focus the present work on this system. 

LS\,5039 is a system with a O6.5V(f) main star and a compact object within a
mildly eccentric $\sim$3.9\,day orbit \citep{2005MNRAS.364..899C,
2011MNRAS.411.1293S}. The source shows periodic X-ray emission with a peak
around the inferior conjunction \citep{2009ApJ...697..592T}, a modulation
practically mirrored by its VHE emission \citep{2005Sci...309..746A,
2006A&A...460..743A}. The X-ray and VHE fluxes have a minimum at orbital phases
$\phi=0.0\mbox{--}0.3$ and a maximum at $\phi=0.6\mbox{--}0.9$, where $\phi=0$
is the periastron phase and the inferior and superior conjunctions take place at
$\phi\approx0.06$ and $\phi\approx0.65$, respectively. 

The observation of LS\,5039 with \emph{Fermi}/LAT starting from 2008 provided a
wealth of new data but also new challenges for the understanding of its nature.
It was shown that LS\,5039 exhibits extremely powerful emission at GeV energies with
a spectral shape consistent with a powerlaw and an exponential cutoff at
energies of a few GeV \citep{2009ApJ...706L..56A}. Long-term observations
confirm that the emission is modulated with the orbital period, with flux
variations of a factor three along the orbit, but the spectral cutoff energy
appears to remain constant \citep{2012ApJ...749...54H}.  The apparent similarity
of the GeV spectrum with that of magnetospheric pulsar emission (hard powerlaw
with an exponential cutoff at a few GeV), prompted the consideration of the
latter as an origin for the gamma-ray emission of LS\,5039
\citep{2009ApJ...706L..56A}. However,  the clear modulation with the orbital
period and the lack of pulsations are hard to reconcile with a magnetospheric
origin. A way to overcome these limitations has been explored in the IC emission
of the unshocked pulsar wind by invoking the striped pulsar wind model
\citep{2011MNRAS.417..532P}, but this approach would result in a
lightcurve modulation amplitude of a factor $\sim$3 larger than the observations
show. \cite{2011MNRAS.418L..49B} considered electron acceleration in the
stellar wind termination shock as a mechanism to obtain a cutoff around a few
GeV, but the stellar wind energetics is hardly enough to explain the GeV
luminosity. In addition, both approaches continue to consider the TeV emission
as arising from a region close to the apex of the contact discontinuity between
the two winds, where pair creation absorption is extremely strong in orbital
phase ranges with clear VHE detections \citep{2008A&A...489L..21B}.

The complex spectral characteristics exhibited by LS\,5039 are not compatible with
the predictions from electromagnetic cascading models, an effect that would
result in a pileup of emitted energy just below the photon-photon pair
production threshold at $\sim$100\,GeV \citep[see][and references
therein]{2008A&A...489L..21B,2010A&A...519A..81C}. 
The combined effects of adiabatic losses, synchrotron and IC cooling, and
anisotropic IC emission applied to a single particle population with a powerlaw
energy distribution are also unable to reproduce a cutoff at a few GeV while
accounting for the hard spectrum from 10\,GeV to 1\,TeV \citep[see also
\citealt{2011A&A...527A...9Z} for LS\,I\,+61\,303]{2008MNRAS.383..467K,
2008A&A...477..691D}. Therefore, present evidence points towards the presence of
at least two particle populations that give rise to the GeV and TeV components
in the high-energy spectrum of LS\,5039. 

In this paper we propose a model that assumes two different locations for
the production of the observed GeV and TeV components of gamma-ray emission. These
two components arise naturally from the orbital motion of a compact pulsar wind
nebula around a massive star. We propose the apex of the contact discontinuity
as the candidate location for the GeV emitter, and a pulsar wind termination
shock in the direction opposite of the star, which appears because of the
orbital motion, as the candidate location for the TeV emitter.

\section{Dynamical model}\label{sec:dyn}

The interaction of the pulsar and stellar winds gives rise to an interface
between the two winds defined by the so-called contact discontinuity (CD), a
surface where the perpendicular velocities of both winds are null. The CD is
bounded on either side by the shocked winds of the pulsar and the star.
Considering a static binary system, the shape of the CD can be analytically
computed by approximating it as the location where the ram pressures of the
winds are in balance \citep[and references therein]{2011ApJ...743....7Z}. In
this scenario, the shape is defined by the ratio of the ram pressures of the
pulsar and stellar winds $\eta_\mathrm{w}\equiv
(\lsd/c)/(\dot{M}v_\mathrm{w})$.  
However, the orbital motion in gamma-ray binaries is bound to modify the shape
of the interaction between the two winds.

Close to the pulsar, the effect of the orbital motion is not significant for the
currently known gamma-ray binaries. At the location of the apex of the CD, the
orbital motion manifests itself in the aberration of the CD, with an angle given
by $\tan{\mu} = v_\mathrm{orb} / v_\mathrm{w}$, where $v_\mathrm{orb}$ is the
orbital velocity and $v_\mathrm{w}$ is the speed of the slowest wind (i.e., the
stellar wind). For the case of LS\,5039, the aberration angle reaches a maximum of
$23^\circ$ at periastron, an angle small enough for the scenario to remain
essentially unchanged from an stationary CD. The location of the apex of the CD,
or wind standoff distance, can be computed from the balance of the ram pressures
of the stellar and pulsar winds as $x_\mathrm{so} =
D/(1+\sqrt{\eta_\mathrm{w}(x_\mathrm{so})})$, where $D$ is the orbital
separation and $\eta_\mathrm{w}(x_\mathrm{so})$ is the pulsar-to-stellar wind
ram pressure ratio at a distance $x_\mathrm{so}$ from the star. When considering
a $\beta$-law for the velocity profile of the stellar wind, as expected from
hot, young stars \citep{1986A&A...164...86P}, $x_\mathrm{so}$ must be obtained
numerically.

Hydrodynamical simulations of stellar and pulsar wind interaction indicate that
the shocked pulsar wind flows roughly parallel to the CD and is accelerated up
to high bulk Lorentz factors within the binary system
\citep{2008MNRAS.387...63B}.  This implies that significant Doppler boosting of
the emission from the pulsar wind shocked in the inner part of the system may be
at play \citep{2008IJMPD..17.1909K,2010A&A...516A..18D}. At larger distances
from the apex, the flow is disturbed by hydrodynamical instabilities and gas
mixing between the two winds becomes significant \citep{2012A&A...544A..59B}.
Observational confirmation of the morphology of the interaction between the
stellar and pulsar winds has been obtained in at least one case of a binary
system known to contain a pulsar. \cite{2011ApJ...732L..10M} resolved the
variable radio nebula of PSR\,B1259$-$63 and deemed it to be compatible with the emission
from cooled electrons originated in the interaction of the stellar and pulsar
winds.

On scales larger than the binary separation the wind interaction region
undergoes dramatic changes with respect to the stationary scenario. The CD is
distorted by the sideways asymmetric interaction with the stellar wind and
results in a bending that would mix the stellar and pulsar winds,
eventually becoming an expanding
disrupted structure interacting with the interstellar medium
\citep{2011A&A...535A..20B}.  Given this orbital effect, the pulsar wind is
expected to be shocked even in the direction opposite from the star, as was
predicted for pulsar binary systems by \cite{2011A&A...535A..20B} and later
confirmed through hydrodynamical simulations in \cite{2012A&A...544A..59B}.
Following \cite{2011A&A...535A..20B} we can obtain an analytical estimate of the
location of the pulsar wind termination shock as the location for which the ram
pressure of the stellar wind owing to Coriolis forces and the ram pressure of
the pulsar wind are in balance, obtaining the following approximate expression
for locations outside of the binary system 
\begin{equation} 
    x_\mathrm{cor}\simeq 
    \sqrt{\frac{\lsd v_\mathrm{w}}{\dot{M}c(2\Omega)^2}}, 
    \label{eq:turnover} 
\end{equation} 
where $\Omega$ is the angular velocity of the compact object around the star.
Owing to Kepler's second law, $\Omega\propto D^{-2}$, where $D$ is the orbital
separation, and therefore $x_\mathrm{cor}\propto D^2$ for eccentric orbits and
constant stellar and pulsar wind properties. Thus, the balance position
$x_\mathrm{cor}$ will be closer to the pulsar at periastron and farther at
apastron.  In spite of the approximate nature of \refeq{turnover}, it is
consistent with the location of the shock in the simulations of
\cite{2012A&A...544A..59B}, and will therefore be used throughout this work. 

The presence and properties of this quasi-perpendicular strong shock at the
Coriolis turnover location sets apart the characteristics of the interaction of
stellar and pulsar wind with respect to that of binary stellar systems. The
non-relativistic stellar wind in binary stellar systems can be smoothly
deflected by sound waves, as found in the simulations of
\cite{2012A&A...546A..60L}. However, sound wave deflection would not affect an
already trans-sonic relativistic pulsar wind, leading to the formation of a
quasi-perpendicular strong shock because of structure bending.  We note that
this approach may not be valid for $\eta_\mathrm{w}\lesssim1.2\times10^{-2}$, as
numerical simulations indicate that a reconfinement shock, with different
characteristics to those explored above, will develop in the pulsar wind for
those cases where the stellar wind vastly overpowers the pulsar wind
\citep{2008MNRAS.387...63B}.

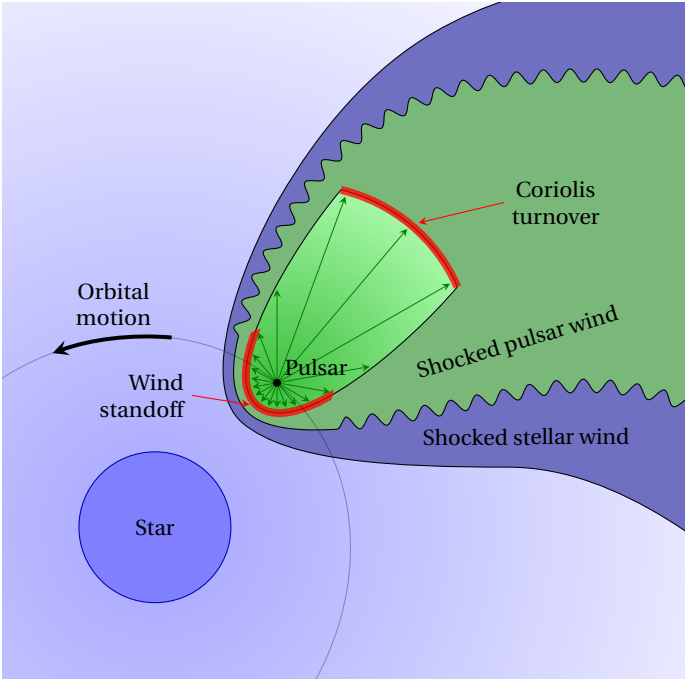
\begin{figure}
    \begin{center}
        \begin{tikzpicture}[scale=1.0,]
\clip (-2,-2) rectangle ($(-2,-2)+(255pt,255pt)$);
\begin{scope}[rotate=50]
\shade[inner color=blue!35,outer color=white] (0,0) circle (8.5);
\filldraw[fill=blue!50,draw=blue!70!black] (0,0) circle (1) node {Star};

\path (2.5,0) coordinate (P);
\path (0,0) coordinate (S);

\path (P)+(-0.4,0) coordinate (apexps);
\path (P)+(2.5,1) coordinate (ps1);
\path (P)+(2.5,-1) coordinate (ps2);

\path (apexps)+(-0.15,0) coordinate (apexcd);
\path (ps1)+(-0.3,+0.3) coordinate (cd1);
\path (ps2)+(-0.4,-1.75) coordinate (cd2);

\path (apexps)+(-0.3,0) coordinate (apexws);
\path (cd1)+(-0.3,+0.25) coordinate (ws1);
\path (cd2)+(-1.0,-0.3) coordinate (ws2);

\path[draw,fill=blue!40!gray!80] (4.5,-7) .. controls +(90:1) and +(-50:1.5) .. (ws2)
        .. controls +(130:1.5) and +(-90:0.8) .. (apexws)
        .. controls +(90:0.8) and +(190:1) .. (ws1) 
        .. controls +(10:1) and +(160:1) .. ++(3,-0.15)
        .. controls +(-20:1.5) and +(110:1.5) .. ++(3,-3)
        .. controls +(-70:3) and +(90:1) .. ++(3,-15);

\path[draw,fill=green!30!gray!80] (6,-6) decorate
[decoration={snake,post=curveto,post length=1.1cm}] 
    { .. controls +(90:1) and +(-40:1.5) .. (cd2)  
    .. controls +(140:1) and +(-90:0.8) .. (apexcd) }
    decorate [decoration={snake,pre=curveto,pre length=1.1cm}] {
    .. controls +(90:0.8) and +(185:1) .. (cd1)
     .. controls +(5:0.5) and +(150:1) .. ++(2.,-0.5)
    .. controls +(-30:1) and +(110:1) .. ++(2.5,-3)
    .. controls +(-70:3) and +(90:1) .. ++(3,-15)};




    \path[draw] decorate [decoration={text along path, text={Shocked pulsar
    wind}}] {($(cd2)+(-0.85,1.35)$) -- ++(3.5,-2.3)};
    \path[draw] decorate [decoration={text along path, text={Shocked stellar
    wind}}] {($(cd2)+(-1.5,0.75)$) -- ++(2.85,-3.35)};

\begin{scope}
    \clip [draw=none] (ps2) .. controls +(180:1) and +(-90:0.7) .. (apexps)
        .. controls +(90:0.7) and +(180:1) .. (ps1) arc (26.565:-26.565:2.23606);
    \shade[inner color=green!50!gray,outer color=green!30] (P) circle (2.7);
\end{scope}
        \path[draw,name path=pws] (ps2) .. controls +(180:1) and +(-90:0.7) .. (apexps)
        .. controls +(90:0.7) and +(180:1) .. (ps1) arc (26.565:-26.565:2.23606);

    \foreach \n in {0,20,...,359} { 
\path[name path=ray] (P) -- +(\n:10);
\draw [line width=0.3pt,shorten >=2pt,green!50!black,-stealth,name intersections={of=pws and ray}] (P) -- (intersection-1) ;
}



\node[black,text width=1.35cm,align=right] (standoff) at (1.15,1.35) {Wind standoff} ;
\draw[red,-stealth,shorten >=3pt] (standoff.east) -- (apexps);

\path (6.25,-0.75) coordinate (coriolis);
\path[name path=foo] (P) -- ($(P)+(10,0)$);
\draw[red,-stealth,shorten >=3pt, name intersections={of=foo and pws}] (coriolis)
node[black,right,text width=1.15cm,text centered] {Coriolis turnover} -- (intersection-1);


\begin{scope}
    \clip (P) circle (0.75);
    \draw[draw=red,line width=3pt,opacity=0.75] (ps2) .. controls +(180:1) and +(-90:0.7) .. (apexps)
            .. controls +(90:0.7) and +(180:1) .. (ps1);
\end{scope}
\draw[draw=red,line width=3pt,opacity=0.75] (ps1) arc (26.565:-26.565:2.23606);

\draw[thin,opacity=0.3] (-1,0) circle (3.5 and 3);
\draw[black,line width=1.5pt,-stealth] ($(S)!1.01!35:(P)$) arc (35:60:3.65) ;
\draw ($(S)!1.05!52:(P)$) node[draw=none,fill=none,above,text width=1.5cm,text
centered] {Orbital motion};

\fill (P) circle (1.5pt) node[above right] {Pulsar};

\end{scope}

\end{tikzpicture}
    \end{center}
    \caption{Sketch of the proposed scenario at the periastron of a close-orbit
        system similar to LS\,5039. The region of the CD where
        significant turbulence, and therefore wind mixing, are expected to take
        place is indicated with a wavy line. The red regions indicate the two
        proposed emitter locations.}
    \label{fig:sketch}
\end{figure}

\section{Location and properties of the non-thermal emitters}

A sketch of the system, as described in the previous section, can be found in
\reffig{sketch}. The influence of the orbital motion on the shape of the pulsar
wind zone gives rise to two distinct regions where most of the energy of the
pulsar wind is deposited in quasi-parallel shocks: the wind standoff and
Coriolis turnover locations. The intermediate regions of the pulsar wind shock,
i.e., those not marked in red in \reffig{sketch}, are not expected to contribute
significantly to the non-thermal emission of the system given the high obliquity
of the impact of the pulsar wind and correspondingly low energy transfer into
non-thermal particles.

The pulsar wind termination region close to the wind standoff location has been
generally considered as the preferred location for acceleration of electrons up
to ultrarelativistic energies, which would give rise to the observed non-thermal
broadband emission through synchrotron and IC processes
\cite[e.g.,][]{2006A&A...456..801D}. However, the intense stellar photon field
at this location, and corresponding high opacities owing to photon-photon pair
production, precludes VHE gamma-rays above $\sim$40\,GeV from reaching the
observer for some orbital phases with clear VHE detections
\citep{2008MNRAS.383..467K}. On the other hand, an emitter with constant
particle injection located at the wind standoff is compatible with the
phenomenology exhibited by the source at GeV energies. The modulation is mainly
driven by the change in the IC interaction angle, with enhanced (reduced)
emission at superior (inferior) conjunction. For the allowed orbital inclination
angles \citep[$20^\circ$--$60^\circ$,][]{2005MNRAS.364..899C}, the modulation is
too strong to reproduce the \emph{Fermi} lightcurve, but this effect may be
mitigated by Doppler boosting. As a proxy for the complex hydrodynamical
properties of the flow \citep{2008MNRAS.387...63B}, we have considered an
effective bulk flow velocity of $0.15c$ in the direction of the axis of symmetry
of the CD apex, which changes along the orbit (see CD aberration in
\refsec{dyn}) but is always close to the radial direction. The bulk flow
velocity of $0.15c$ used here applies exclusively to the Doppler effect averaged
over the CD cone and with respect to the observer line-of-sight. Therefore, it
is still possible, and, given the simulations of \cite{2008MNRAS.387...63B},
probable, that the intrinsic bulk flow velocity of the shocked pulsar wind is
higher.

The pulsar wind termination shock at the Coriolis turnover location is a good
candidate for the VHE emitter. The reduced stellar photon field density implies
pair production absorption characteristics consistent with the observed spectra
at VHE. For scenarios in which the stellar wind ram pressure is dominant, the
pulsar wind solid angle subtended by the Coriolis turnover shock is much lower
than that of the apex region. This is consistent with the reduced emission power
of the TeV component with respect to the GeV component.

The magnetic field of the shocked pulsar wind is generally well known in
plerions from observations \citep[e.g.,][]{1984ApJ...283..710K}, but the
so-called sigma problem precludes one from deriving it theoretically.  The
situation in gamma-ray binaries is even more complex, as there might be a
reduction of the post shock magnetic field strength owing to the reacceleration
of the flow \citep{2012MNRAS.419.3426B}. Since knowledge of the post-shock
magnetic field could only be obtained via a magnetohydrodynamical simulation of
the system, we have chosen to parametrize the magnetic field energy density as a
fraction $\xi$ of the pre-shock equipartition magnetic field. We define the
latter at the balance of the magnetic field energy density and the kinetic
energy density of the pulsar wind at the shock. Therefore, for a constant value
of $\xi$, the post-shock magnetic field strength will behave along the orbit as
$B\propto1/r_\mathrm{p}$, where $r_\mathrm{p}$ is the distance of the shock to
the pulsar. 

\begin{figure}[t]
    \includegraphics[width=256pt]{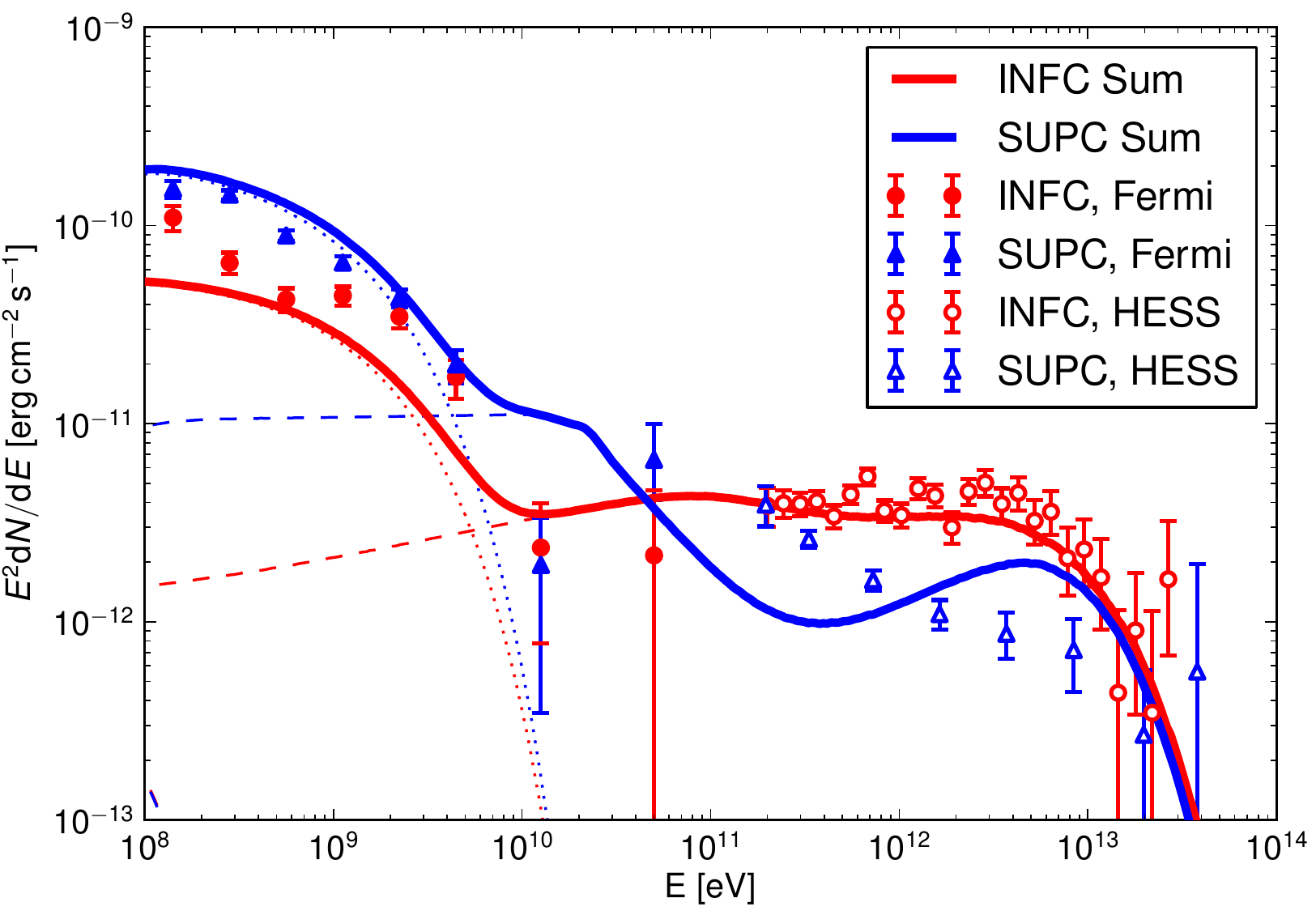}
    \caption{Spectral energy distribution of LS\,5039. The computed emission and
        observational data during the  inferior conjunction ($0.45<\phi<0.9$) is
        shown in red and during the superior conjunction ($0.9<\phi<0.45$) in blue.
        The emission components from the wind standoff and Coriolis turnover locations
        are indicated with a dotted and dashed line, respectively. }
    \label{fig:sed}
\end{figure}

\subsection{Maximum electron energy}\label{sec:maxener}

A distinct feature of the GeV and TeV spectra of LS\,5039 are the cutoffs present at
a few GeV and above 10\,TeV, respectively. The former is apparently constant
along the orbit, whereas owing to the softer TeV spectrum during superior
conjunction the latter has only been detected clearly during inferior
conjunction. These spectral cutoffs can be directly related to the maximum
energy of the electrons in the respective particle populations. The maximum
energy that can be reached in any acceleration process can be determined by the
balance of the acceleration rate with the energy losses owing to radiative
processes. The acceleration timescale  is characterized as
$t_\mathrm{acc}=\eta_\mathrm{acc} r_\mathrm{L}/c$, where $r_\mathrm{L}$ is the
Larmor radius, and $\eta_\mathrm{acc}\geq1$ parametrizes the efficiency of the
acceleration process (see \refsec{eta} below for a discussion of
$\eta_\mathrm{acc}$). For electrons accelerated close to the binary system, the
dominant radiative losses will be related to synchrotron and IC emission. A study
of LS\,5039 led \cite{2008MNRAS.383..467K} to the conclusion that the production of
the HESS spectrum required $\eta_\mathrm{acc}\lesssim10$ for emitters close to
the binary system. The wind standoff location will present both higher seed
photon density and magnetic field, owing to its proximity to the star and the
denser preshock pulsar wind, and the higher energy losses will lead to a lower
maximum electron energy compared to the location outside the binary system. 

The highest energy particles in the Coriolis turnover location may be used to
gain information on its magnetic field. The maximum electron energies obtained
under dominant synchrotron or IC losses in the Klein-Nishina regime depend
differently on the magnetic field: $E_\mathrm{e,max}\mathrm{(syn)}\propto
B^{-1/2}$, and $E_\mathrm{e,max}\mathrm{(IC,KN)}\propto B^{3.3}$
\citep{2008MNRAS.383..467K}.  For a given $\eta_\mathrm{acc}$, the highest
maximum energies will be obtained for a narrow range of magnetic field strengths
where synchrotron losses are only slightly dominant over IC losses. 

\subsection{Emission model}\label{sec:model}

As a first approximation, we will simplify the scenario mentioned above by
considering the two emitting regions as point-like, homogeneous emitters located
at the wind standoff and Coriolis turnover locations. They are located on the
axis of symmetry of the system, and corotate with the pulsar around the star.

At each of the two positions, we consider an injection in accelerated electrons
with a total luminosity as a fraction of the pulsar spindown luminosity, which
may be different for the two locations but constant along the orbit. Given the
short cooling timescales with respect to the dynamical timescales of the system,
a steady state approach to computing the evolved particle spectrum can be taken
\citep{2007MNRAS.380..320K, 2011A&A...527A...9Z}. Radiative losses through
synchrotron and IC cooling are considered, as well as non-radiative adiabatic
losses. Adiabatic expansion losses can be approximated as $t_\mathrm{ad}\simeq
R/v_\mathrm{exp}$, where $R$ is the characteristic size of the expanding emitter
and $v_\mathrm{exp}$ the expansion velocity. For our two emitters, we have
approximated the characteristic size as the distance to the pulsar and the
expansion velocity as $c/3$. Nevertheless, under these approximations adiabatic
losses are not dominant in any electron energy range. 

The resulting emission spectra is then computed for each of the two positions.
The anisotropic nature of IC is of utmost relevance in gamma-ray binaries
\citep{2008MNRAS.383..467K,2008A&A...477..691D}, and is taken into account using
the simple analytical presentation of the differential cross-sections for
anisotropic inverse Compton scattering of \cite{1981Ap&SS..79..321A} and
considering the black body radiation from the star as the seed photon field. The
pair production absorption along the line of sight to the observer is then
computed and applied to the intrinsic spectrum.  To compare with spectral
observations spanning significant fractions of the orbit, we computed the
average of the emission along the observation period from a sampling of 100
points along the orbit.

\section{Application to LS\,5039}\label{sec:ls}

\begin{figure}[t]
    \includegraphics[width=256pt]{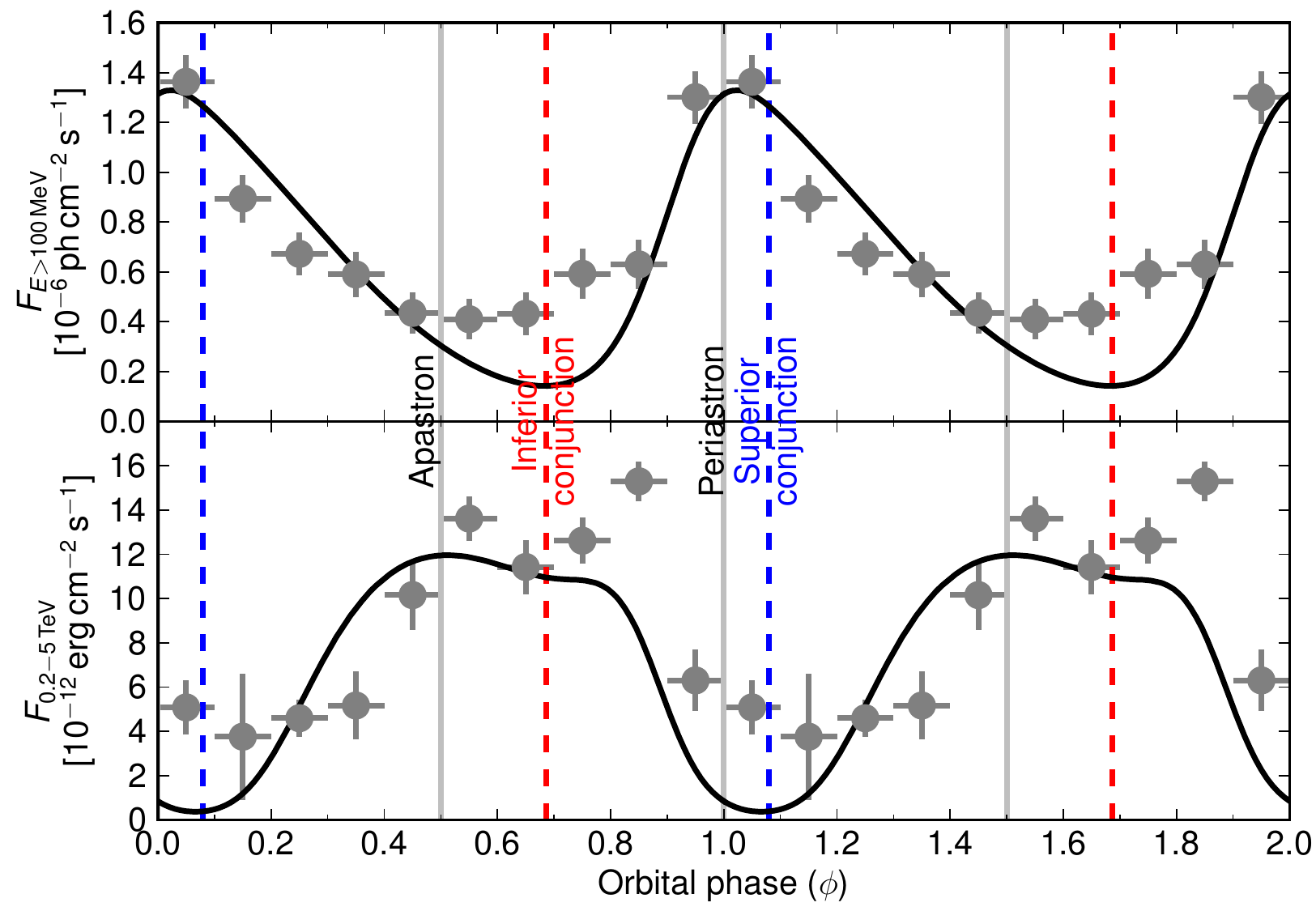}
    \caption{\emph{Top:} \emph{Fermi}/LAT orbital lightcurve of LS\,5039
        in the energy range 100\,MeV to 30\,GeV, along with the
        computed gamma-ray emission. \emph{Bottom:} HESS orbital lightcurve in the
        range 0.2 to 5\,TeV. Two orbital periods are shown for clarity.}
    \label{fig:lc}
\end{figure}


The stellar mass-loss rate and pulsar spin-down luminosity of LS\,5039 have been
constrained into a relatively narrow range by a variety of studies of the
system. The lack of orbital variability of the photoelectric soft X-ray
absorption places an upper limit on the stellar mass-loss rate of
$5\times10^{-8}\,\msolyr$ for a point-like X-ray emitter \citep{2007A&A...473..545B}
and of $1.5\times10^{-7}\,\msolyr$ for an extended emitter, as expected in a PWS
scenario \citep{2011MNRAS.411..193S}. 
Note, however, that for these calculations an emitter located at the position of
the compact object was assumed.  If the X-ray emitter is located outside the
orbit of the compact object, in a position similar to the Coriolis turnover
location mentioned above, these constrains would likely be relaxed.
A direct measure through H$\alpha$ line
measurements yields 3.7 to $4.8\times10^{-7}\,\msolyr$
\citep{2011MNRAS.411.1293S}, but this method could overpredict mass-loss rates
by a factor of two if there is clumping present in the stellar wind
\citep{2004A&A...413..693M}. The non-detection of thermal X-ray emission from
the shocked stellar wind in the X-ray spectra of LS\,5039 places a robust upper
limit on the  pulsar spin-down luminosity of $\sim6\times10^{36}\,\ergs$
\citep{2011ApJ...743....7Z}.

In the application of the model to LS\,5039 we considered a stellar wind with a
mass-loss rate of $\dot{M}=10^{-7}\,\msolyr$, a terminal wind velocity of
$v_\mathrm{w}=2400\,\kms$, and a pulsar spin-down luminosity of
$L_\mathrm{sd}=5\times10^{36}\,\ergs$, which results in a ram pressure ratio of
$\eta_\mathrm{w,\infty}\simeq0.1$.  Under these conditions the distance of the
wind standoff emitter to the star varies between half and two thirds of the
orbital separation for periastron and apastron, respectively. In order to match
the powerful observed GeV emission, a non-thermal electron injection power of
$3.5\times10^{35}\,\ergs$ is required at this location, corresponding to $7\%$
of the pulsar spin-down luminosity. In spite of the strong radiative losses
owing to the intense stellar photon field, a relatively low acceleration
efficiency of $\eta_\mathrm{acc}\sim2000$ is required to reproduce the spectral
cutoff at a few GeV. The spectral shape of the GeV component is qualitatively
well reproduced during the inferior and superior conjunctions (\reffig{sed}),
and its orbital lightcurve agrees well with the observational data
(\reffig{lc}). The computed emission is lower than the observed one only during
inferior conjunction, which could be related to uncertainties in the Doppler
boosting direction. While here we have considered a single quasi-radial
direction for the entire emission from the shocked pulsar wind, a wider spread
in the direction of the different parts of the post-shock flow \citep[as seen in
the simulation results of][]{2012A&A...544A..59B} could mean that certain parts
are boosted in a direction closer to the observer line-of-sight, which would
lead to increased emission.

The emitter located at the Coriolis turnover region requires a lower injection
energy of $5\times10^{34}\,\ergs$, or $1\%$ of the pulsar spindown luminosity,
which is consistent with the lower solid angle subtended by this region. The
lower magnetic field and larger distance to the star, resulting in lower
synchrotron and IC losses, favor the acceleration of particles up to very high
energies at this location. However, a very efficient accelerator, with
$\eta_\mathrm{acc}<20$, is still required to obtain the multi TeV electrons that
give rise to the gamma-ray emission up to $\sim$10\,TeV detected by HESS. The
requirement for such energetic particles also constrains the magnetic field in
the emitter (see \refsec{maxener}), to a fraction $\xi=0.025$ of the
equipartition preshock magnetic field, resulting in magnetic fields that vary
along the orbit in the range from 0.2 to 0.4\,G.  The computed emission provides
a very good spectral match to observed data during inferior conjunction, with a
flat spectrum up to the spectral cutoff around 10\,TeV, as can be seen in
\reffig{sed}. At superior conjunction the computed spectra has the
characteristic shape of pair-production absorption, with a minimum around a few
hundred GeV, whereas the observed spectrum is closer to a powerlaw.
Additionally, the overall flux around periastron and superior conjunction is
underestimated, as seen in \reffig{lc}, bottom. Possible reasons for this
discrepancy are discussed below in \refsec{tevsupc}. Under the assumption of a
constant $\xi=0.025$ for all pulsar wind shocks, the resulting magnetic
field of the wind standoff postshock material is between 0.6 and 0.9\,G.

The inclination of the orbital system has a strong effect on the orbital
modulation of the IC emission, both because of the emission angular dependency
(affecting both the HE and VHE light curves) and because of the variation in
pair-production opacity along the orbit with inclination (affecting the VHE
light curve).  We found that the value of the inclination angle that best
leverages the influence of these two effects in reproducing the HE and VHE gamma
ray light curves is $i=45^\circ$. Taking the mass function
$f(m)\approx0.0049\pm0.0006\,\msol$ measured by \cite{2011MNRAS.411.1293S} and a
primary star mass of $22.6^{+3.4}_{-2.6}\,\msol$, this inclination results in a
lower limit to the mass of the compact object of $\sim2\,\msol$. 

\section{Discussion}\label{sec:disc}

\subsection{Acceleration efficiencies}\label{sec:eta}

In \refsec{ls} we showed that significantly different acceleration efficiencies
are required in our model for the two emitter locations. It is beyond the scope
of this paper to study the specific acceleration mechanisms present in the PWS,
but a few insights can be gained through the acceleration efficiency
requirements of the two locations. 

Given the ultrarelativistic velocity of the pulsar wind, the pulsar wind shocks
will necessarily be relativistic. The detailed physics of realistic particle
acceleration in astrophysical relativistic shocks is currently not well
characterized, even though significant progress has been made in the last decade
\citep[see, e.g.,][for a review]{2012AIPC.1439..194L}. One of the key points
that has resulted from these developments is that particle acceleration
efficiency, and therefore electron maximum energy, depends strongly on magnetic
turbulence around the shock and on the size of the shock.  The different
acceleration efficiencies at the wind standoff and Coriolis turnover could be
considered an indication of the level of magnetic turbulence present around
the relativistic shocks, and in particular in the postshock regions.  The lower
density and the influence of turbulent flow from the outer regions of the CD may
give rise to a significantly higher magnetic field inhomogeneity in the Coriolis
turnover postshock material with respect to the wind standoff location.  In
addition, the two shocks have significantly different sizes, with characteristic
dimensions of $\sim0.1D$ and $\sim5D$ for the wind standoff and Coriolis
turnover shocks, respectively. The shock size is key in determining how long a
particle can be retained close to the shock, and therefore how much energy can
be transfered to it.  These two effects, magnetic tubulence and shock size,
could work together and result in the required $\eta_\mathrm{acc}$ of $\sim$20
and $\sim$2000 for the Coriolis turnover and wind standoff locations,
respectively.

\subsection{X-ray emission}

The model presented here does not aim to explain the whole broadband spectrum of
gamma-ray binaries, but is limited to the HE and VHE gamma-ray bands. However,
the X-ray emission of LS\,5039 deserves discussion owing to the resemblance of its
orbital modulation to that of VHE emission, as well as the significant fraction
of energetic output of the system it represents. \cite{2009ApJ...697..592T}
proposed a common origin of the X-ray and VHE emission, the orbital modulation
of both arising from dominant adiabatic losses. In order to obtain the observed
high X-ray fluxes through synchrotron emission, the emitter magnetic field was
required to be on the order of 3\,G. In \refsec{ls} we have shown, however, that
to reproduce the observed TeV spectrum from the Coriolis turnover location a
magnetic field on the order of 0.2 to 0.4\,G is required. A stronger magnetic
field would lead to a dominance of synchrotron cooling over Klein-Nishina IC for
the highest energy particles and, as a result, a softening of the spectrum above
100\,GeV that would be in contradiction with the observed inferior conjunction
spectrum (see \reffig{sed}).  Even considering dominant adiabatic losses, as
done by \cite{2009ApJ...697..592T}, results in a spectrum above 5\,TeV that is
too soft compared to the observations.  Nevertheless, it is clear that a
magnetic field strength of 0.2 to 0.4\,G is insufficient to produce the observed
X-ray fluxes.  The 0.3 to 10\,keV X-ray flux observed by Suzaku varies in the
range $5\mbox{--}12\times10^{-12}\,\ergcms$, whereas the fluxes computed through
the proposed model are more than an order of magnitude fainter at around
$3\times10^{-13}\,\ergcms$.

The two magnetic field requirements, however, need not be in contradiction. The
turbulence present in the postshock flow at the Coriolis turnover location
results in strong fluctuations in the magnetic field.  Close to the shock,
these could be lower and result in effective magnetic field strengths of
$\sim$0.5\,G, as required to account for the observed TeV emission. As
particles flow downstream, they might encounter stronger magnetic field
turbulence with local magnetic field strengths on the order of or higher than
3\,G, resulting in stronger synchrotron emission. However, the process
through which this emission may be modulated along the orbit is still unclear.

The electron population accelerated at the wind standoff location does not
contribute significantly to the X-ray emission of the system. Even though the
magnetic field at this location is higher (between 0.6 and 0.9\,G) owing to its
proximity to the pulsar, electrons are not energetic enough to radiate
synchrotron in the X-ray band, displaying a spectral cutoff at around 100\,eV.
In addition, because of the higher stellar photon field density, the IC emission
channel is more efficient than the synchrotron, resulting in a low
synchrotron-to-IC-luminosity ratio.

\subsection{TeV emission at superior conjunction} \label{sec:tevsupc}

The most significant deviation between the computed fluxes of the proposed model
and the observed phenomenology of the source is the behavior of TeV emission
around superior conjunction. As seen in \refsec{ls}, the predicted integrated
flux above 200\,GeV is below the observed flux (\reffig{lc}) and the predicted
spectrum shape presents the characteristic pair-production absorption features
instead of a powerlaw spectrum (\reffig{sed}). 

A simple reason for this discrepancy could be our simplification of the emitter
as a point-source instead of as an extended source where part of the emitter
would be above the orbital plane and could be significantly less absorbed
\citep[see, e.g.,][for a study of the optical depth of various locations around
the binary system]{2008MNRAS.383..467K}. It must be noted that the Coriolis
turnover shock would be compressed in the tangential direction by Coriolis
forces, but not in the direction perpendicular to the orbital plane. Therefore,
it would rise above the plane following an opening angle of
$\theta\simeq28.6^\circ(4-\eta_\mathrm{w}^{2/5})\eta_\mathrm{w}^{1/3}$
\citep{2008MNRAS.387...63B}, resulting in heights between 2 and
$5\times10^{12}$\,cm above the orbital plane at periastron and apastron,
respectively.  Additionally, the energy absorbed through pair production is
bound to be reemitted by secondary electrons \citep{2008A&A...482..397B}, and,
for certain values of magnetic field and photon density, by efficient pair
cascades \citep[e.g.,][]{2006JPhCS..39..408A, 2007A&A...464..259B,
2010A&A...519A..81C}. However, in an electromagnetic cascade scenario a
significant part of the absorbed energy would also end up being emitted as
$\sim$10\,GeV photons, just below the threshold for $\gamma\gamma$ pair
production on stellar photons, an excess that would be in contradiction with the
relatively low flux spectral point detected by \emph{Fermi}/LAT at
$\sim$10\,GeV. 

\subsection{Application to other gamma-ray binaries}

LS\,5039 is an ideal system to test wind interaction models because of its radial,
well-determined stellar wind. The rest of the currently known gamma-ray binaries
present a harder challenge either because the stellar wind structure is
significantly more complex or because our current phenomenological knowledge of
them is rather limited. The former applies to PSR\,B1259$-$63, LS\,I\,+61\,303 and HESS\,J0632+057, which
contain a Be star with a dense equatorial near-Keplerian disk, whereas the
latter is the case for 1FGL\,J1018.6$-$5856. The extremely large orbital separation of PSR\,B1259$-$63\
sets it apart from the other gamma-ray binaries, as radiative processes in the
wind interaction region are only significant during a short period of time
around periastron. During this time, the interaction of the pulsar wind with the
stellar equatorial disk gives rise to broadband non-thermal emission with a rich
phenomenology \citep[e.g.,][]{2005A&A...442....1A, 2009MNRAS.397.2123C,
2011ApJ...736L..11A}. Given the presence of the dense equatorial disk during the
compact object passage, the model presented here is not applicable.  For the
other two gamma-ray binaries with Be stars, LS\,I\,+61\,303 and HESS\,J0632+057, the situation is
different as most of their emission takes place away from periastron and
therefore far from the equatorial disk. Additionally, the compact orbit of LS\,I\,+61\,303
means that the stellar disk is probably disrupted by the compact object passage
\citep{2007A&A...474...15R}. These two systems have a wider orbit than LS\,5039, and
their radial component of the stellar wind is less powerful, resulting in a
Coriolis turnover location farther from the system. Our model, however, provides
a plausible explanation for the orbit-to-orbit instability of VHE and X-ray
emission from LS\,I\,+61\,303. The wind standoff region is quite stable under wind
irregularities, so that GeV emission is expected to be more stable, but the
shock properties at the Coriolis turnover location are much more dependent on
the turbulence generated along the CD, which is in turn strongly affected by
wind irregularities. This could mean that for certain wind interaction regimes
this shock might even disappear, leading to the VHE/X-ray instability of LS\,I\,+61\,303.

1FGL\,J1018.6$-$5856 is the only other gamma-ray binary, along with LS\,5039, to
host an O-type star.  Its similar-sized orbit makes it also a good candidate to
apply our model and further our understanding of the source. The HE gamma-ray
spectrum indeed agrees well with the prediction from a wind standoff location
emitter \citep{2012Sci...335..189F}. However, its detection at VHE is still not
clear, as it is in confusion with a nearby extended source and no flux
variability has been detected yet \citep{2012A&A...541A...5H}. In addition, the
orbital parameters of the system are currently unknown, making any theoretical
predictions highly uncertain. However, the general observational features appear
to be similar to LS\,5039 and as such could be consistent with the framework
presented by our model.

\section{Conclusion}\label{sec:conclusion}

The results presented here show that the two-emitter model for the GeV-TeV
emission of LS\,5039 provides a new explanation to the non-trivial problem of the
origin of the HE and VHE spectral components of gamma-ray binaries. These two
locations arise naturally from the interaction of the stellar and pulsar winds
in an orbiting system.

However, the development of this model and comparison with the increasingly
better-quality observations at GeV and TeV showcases the intrinsic difficulty in
understanding the processes responsible for non-thermal emission in gamma-ray
binaries. The compound effects of acceleration, (magneto)hydrodynamical, and
radiation processes are impossible to disentangle and require a comprehensive
approach to the problem. Even though the best method to unveil the workings of
gamma-ray binaries would be radiation-coupled magnetohydrodynamical simulations,
the large amount of unknown quantities that would be required for such
simulations still make simpler approaches, like the one presented in this paper,
a safer, more confident way to advance in our understanding.

\begin{acknowledgements}
The authors thank the referee, Guillaume Dubus, for his thorough review and
helpful comments on the manuscript. 
We thank D.~Hadasch for providing the \emph{Fermi}/LAT observational data.
V.Z.~acknowledges support by the Generalitat de Catalunya through the Beatriu
de Pin\'os program, the Max-Planck-Gesellschaft, and the Spanish MICINN under
grant AYA\-2010-21782-C03-01.  V.B.-R.~acknowledges support by the Spanish
Ministerio de Ciencia e Innovaci\'on (MICINN) under grants AYA2010-21782-C03-01
and FPA2010-22056-C06-02.  V.B.-R.~acknowledges financial support from MINECO
through a Ram\'on y Cajal fellowship.
\end{acknowledgements}

\balance


\end{document}